\documentclass{ws-ijmpe}

\def\Rslash{\slash \!\!\!\! R}
\def\RMSSM{\Rslash{\rm MSSM}}
\def\GeV{\hbox{\rm GeV}}
\def\eV{\hbox{\rm eV}}

\def\superhu{\hat H_u}
\def\superhd{\hat H_d}
\def\superq{\hat Q}
\def\superu{\hat U^c}
\def\superd{\hat D^c}
\def\superl{\hat L}
\def\supere{\hat E^c}

\def\tildehu{\tilde H_u}
\def\tildehd{\tilde H_d}

\begin{document}

\markboth
{M. G\'o\'zd\'z and W. A. Kami\'nski} 
{Neutralino Induced Majorana Neutrino Transition Magnetic Moments}

%
%

\title{NEUTRALINO INDUCED MAJORANA NEUTRINO\\TRANSITION MAGNETIC MOMENTS}

\author{MAREK G\'O\'ZD\'Z and WIES{\L}AW A. KAMI\'NSKI}

\address{
Department of Informatics, Maria Curie-Sk{\l}odowska University, \\
pl. Marii Curie--Sk{\l}odowskiej 5, 20-031 Lublin, Poland \\
mgozdz@kft.umcs.lublin.pl, kaminski@neuron.umcs.lublin.pl
}

\maketitle

\begin{history}                 %
\end{history}                  	%

\begin{abstract}
  We calculate the effect of neutrino-neutralino mixing on the neutrino
  magnetic moment and compare it with the contribution of pure
  particle-sparticle loop. We have found that the dominated mechanism is
  still the bare loop, and that the bilinear insertions on the external
  neutrino lines contribute at least one order of magnitude weaker.
\end{abstract}

\section{Bilinear $R$-parity breaking in MSSM and neutrino magnetic
  moment from RpV loops}

Supersymmetry is believed to be the necessary ingredient in the high
energy regime of particle physics, and that it should be included in all
models of grand unified theories (GUT). Despite the fact that it has
been formulated long before the string theory, it soon turned out that
the latter needs supersymmetry to include not only interactions,
mediated by bosonic strings, but also matter in the form of fermionic
strings. Since the string and M-theories suffer from great complexity
and infinit number of solutions, simpler phenomenological models are
used, from which one of the most popular is the minimal supersymmetric
standard model (MSSM).\cite{mgozdz:mssm}

In its basic formulation the MSSM exhibits the same feature as the
ordinary standard model regarding the conservation of the lepton and
baryon numbers. This feature is realized by an artificial introduction
of the so-called $R$-parity defined as a~multiplicative quantum number
assigned to the particles according to the formula $R = (-1)^{3B+L+2S}$,
where $B$, $L$, and $S$ are the baryon, lepton, and spin numbers,
respectively. In its more general form, however, the MSSM does not
preserve the $R$-parity either through a~spontaneous symmetry
breaking\cite{mgozdz:1RpV} or by explicit retaining the previously
rejected $R$-parity violating (RpV) terms\cite{mgozdz:2RpV,mgozdz:3RpV}
in the superpotential. In such a~case the basic superpotential
\begin{equation}
  \label{mgozdz:wmssm}
    W =
      \epsilon_{ab} \left [ 
      (\mathbf{Y}_E)_{ij} \superl_i^a \superhd^b \supere_j
    + (\mathbf{Y}_D)_{ij} \superq_i^{ax} \superhd^b \superd_{jx}
    + (\mathbf{Y}_U)_{ij} \superq_i^{ax} \superhu^b \superu_{jx}
    + \mu \superhd^a \superhu^b \right ] \nonumber
\end{equation}
is extended by 
\begin{equation}
  \label{mgozdz:rpvmssm}
    W_{\Rslash} = \epsilon_{ab} \left [
  \lambda_{ijk} \superl_i^a \superl_j^b \supere_k + \lambda_{ijk}'
  \superl_i^a \superq_j^b \superd_k \right ] + \lambda_{ijk}''
  \superu_i\superd_j\superd_k + \epsilon_{ab} \kappa_j\superl_j^a\superhu^b,
\end{equation}
where the superfields are denoted by a~tilde. The quark and lepton
left-handed doublets are $Q$ and $L$, and the right-handed singlets are
$U$, $D$, and $E$, respectively.

The $\lambda$, $\lambda'$, and $\kappa_i$ coupling constants mediate
lepton number violating interactions. The $\lambda''$ breaks the baryon
number and should be treated carefully because of the very stringent
limits on the proton life-time; it is therefore often set to zero to
avoid unnecessary problems.

For completeness the $\RMSSM$ Lagrangian is supplied with mass terms and
the mechanism of supergravity (SUGRA), in which the supersymmetry is
broken at the Planck scale by explicit soft terms. (See
Refs.~\cite{mgozdz:mssm,mgozdz:theory} for details.)

In this paper we neglect all the trilinear couplings and concentrate on
the bilinear $\kappa$'s. The most important implication of the existence
of the bilinear RpV terms is mixing between different types of
particles: neutrinos with neutralinos
$(\nu_{1,2,3},\tildehu^0,\tildehd^0,\tilde B^0,\tilde W^3)$, as well as
charged leptons with charginos, and neutral Higgs bosons with
sneutrinos.
%
\begin{figure}[t]
  \centerline{\psfig{file=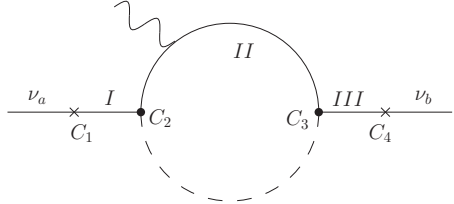,width=0.5\textwidth}}
  \caption{\label{mgozdz:diagram} Feynman diagram with
    neutrino-neutralino mixing on the external lines, leading to the
    Majorana neutrino transition magnetic moment.}
\end{figure}

The $\RMSSM$ provides an elegant mechanism of generating Majorana
neutrino mass terms. It is done by considering RpV particle-sparticle
loops (mostly lepton-slepton or quark-squark) without the need of
introducing a~heavy right-handed singlet neutrino, ie. it provides an
attractive alternative to the popular see-saw mechanism. By introducing
an interaction vertex between one of the virtual particles and an
external photon, a~Majorana neutrino transition magnetic moment is
generated. These processes have been described
already\cite{mgozdz:nu_mm1} but the previous calculations were based on
many simplifying assumptions. The present work is a~continuation of
a~series of papers\cite{mgozdz:nu_mm2,mgozdz:theory} in which we present
exact analytic formulas, and perform the numerical calculations using
GUT constrained $\RMSSM$. The GUT conditions give the starting point for
the renormalization group equations (RGE) from which the low energy
spectrum of the model is obtained. This approach has been recognized as
the most exact by now. In this conference contribution we present
calculations of the transition magnetic moment generated in a~mechanism
in which the external neutrino lines contain bilinear insertions,
causing them to transform into neutralinos.\cite{mgozdz:JDV} The
relevant Feynman diagram is depicted on Fig.~\ref{mgozdz:diagram}. There
is eleven possible configurations, presented in Tab.~\ref{mgozdz:tab1},
which all contribute to the magnetic moment.
%
\begin{table}[!t]
\tbl{\label{mgozdz:tab1}All possible loops with neutrino-neutralino
  mixing on the external lines. The names of the columns correspond to
  the denotations on Fig.~\protect\ref{mgozdz:diagram}.}{
  \begin{tabular}{ccccccc}
    \hline\hline
    $I$ & $II$ & $III$ & $C_1$ & $C_2$ & $C_3$ & $C_4$ \\
    \hline
    $\tilde H_u$ & $u\tilde u$ & $\tilde H_u$ & 
    $\kappa_a$ & $\sqrt{2} m_u/v_u$ & $\sqrt{2} m_u/v_u$ & $\kappa_b$ \\
    $\tilde H_u$ & $u\tilde u$ & $\tilde B$ &
    $\kappa_a$ & $\sqrt{2} m_u/v_u$ & $-g'/(3\sqrt{2})$ & $g'\omega_b$ \\
    $\tilde H_u$ & $u\tilde u$ & $\tilde W^3$ &
    $\kappa_a$ & $\sqrt{2} m_u/v_u$ & $-g/\sqrt{2}$ & $g\omega_b$ \\
    $\tilde B$   & $q\tilde q$ & $\tilde B$ &
    $g'\omega_a$ & $-g'/(3\sqrt{2})$ & $-g'/(3\sqrt{2})$ & $g'\omega_b$ \\
    $\tilde B$   & $l\tilde l$ & $\tilde B$ &
    $g'\omega_a$ & $-g'/\sqrt{2}$ & $-g'/\sqrt{2}$ & $g'\omega_b$ \\
    $\tilde W^3$ & $u\tilde u$ & $\tilde W^3$ &
    $g\omega_a$ & $-g/\sqrt{2}$ & $-g/\sqrt{2}$ & $g\omega_b$ \\
    $\tilde W^3$ & $d\tilde d$ & $\tilde W^3$ & 
    $g\omega_a$ & $g/\sqrt{2}$ & $g/\sqrt{2}$ & $g\omega_b$ \\
    $\tilde W^3$ & $l\tilde l$ & $\tilde W^3$ &
    $g\omega_a$ & $g/\sqrt{2}$ & $g/\sqrt{2}$ & $g\omega_b$ \\
    $\tilde B$   & $u\tilde u$ & $\tilde W^3$ &
    $g'\omega_a$ & $-g'/(3\sqrt{2})$ & $-g/\sqrt{2}$ & $g\omega_b$ \\
    $\tilde B$   & $d\tilde d$ & $\tilde W^3$ &
    $g'\omega_a$ & $-g'/(3\sqrt{2})$ & $g/\sqrt{2}$ & $g\omega_b$ \\
    $\tilde B$   & $l\tilde l$ & $\tilde W^3$ &
    $g'\omega_a$ & $g'/\sqrt{2}$ & $g/\sqrt{2}$ & $g\omega_b$ \\
    \hline\hline
  \end{tabular}}
\end{table}
%
The contribution to the Majorana neutrino magnetic moment from the
discussed diagrams is given by (in Bohr magnetons $\mu_B$)
\begin{equation}
  \mu_{ab} = (1-\delta_{ab}) \frac{m_{e^1}}{4\pi^2}
  \ \left(C_{1a}
    \frac{C_2 C_3}{m_{\chi_I}m_{\chi_{III}}}
    C_{4b} \right ) \ \sum_{jk}
  \left [
    3\frac{w_{jk}^{(q)}}{m_{q^j}} Q_{q^j} +
    \frac{w_{jk}^{(\ell)}}{m_{\ell^j}} Q_{\ell^j}
  \right ] \mu_B.
\label{mgozdz:mu}
\end{equation}
Here we have denoted the sneutrinos' vacuum expectation values by
$\omega$, the electric charge of a~particle (in units of $e$) by $Q$.
The dimensionless loop functions $w$ take the forms
\begin{equation}
  w_{j,k}^{(q)} = \frac12 \sin(2\theta^k) \left[
    \frac{x_2^{jk}\log(x_2^{jk}) - x_2^{jk} +1}{(1-x_2^{jk})^2} -
    (x_2 \to x_1) \right ],
\label{mgozdz:w}
\end{equation}
where $\theta$ is the squarks' mixing angle and $x_i^{jk}=(m_{q^j}
/ m_{\tilde q_i^k})^2$. A~similar expression holds for
$w_{j,k}^{(\ell)}$ with (s)quarks replaced by (s)leptons. The sum over
$j$ and $k$ in Eq.~(\ref{mgozdz:mu}) accounts for all the possible
quark-squark and lepton-slepton configurations for given
neutralinos. The factor 3 in front of $w^{(q)}$ counts the three quark
colors.

\section{Numerical calculations}

In order to reduce the great number of free parameters of the model we
assume the GUT scenario at high energies.\cite{mgozdz:theory} The
initial values of the $\kappa$'s are drawn randomly and evolved down
using RGE equations. Then the neutrino mass matrix is calculated and
compared with the mass matrix obtained from experimental data. If in
agreement, the magnetic moment is calculated.

The phenomenological neutrino mass matrices can be calculated when one
knows the three neutrino mixing angles and the two differences of masses
squared. The best fit values of these parameters
are:\cite{mgozdz:nu-mass}
\begin{eqnarray}
  |m_1^2 - m_2^2| = 7.1 \times 10^{-5} \eV^2, \qquad \qquad
  |m_2^2 - m_3^2| = 2.1 \times 10^{-3} \eV^2, \\
  \sin^2(\theta_{12}) = 0.2857, \qquad
  \sin^2(\theta_{23}) = 0.5,    \quad
  \sin^2(\theta_{13}) = 0.
\label{mgozdz:bestfit}
\end{eqnarray}
Additionally, it is necessary to know at least one element of the mass
matrix. For example one can assume a~certain scenario of the hierarchy
of masses and set the lowest mass to zero. The normal hierarchy aligns
the masses as $0=m_1 < m_2 \ll m_3$ whence in the inverted hierarchy
$0=m_3 \ll m_1 \le m_2$. 
A~separate problem arises from the unknown CP phases which enter the
Majorana neutrino mass matrix. We coupe with this problem twofold. The
simpler case is to assume that the CP symmetry is conserved and neglect
the phases. The more complicated approach requires to check all possible
values of the phase factors for each of the mass matrix elements
separately and for each element pick the combination which leads to its
highest value. In such a~case the obtained matrix do not correspond to
any physically allowed situation, but instead presents an upper bound on
all its elements. It it therefore useful in discussion of the allowed
parameter space of the model. We call the such obtained matrices `maximal'.
%
\begin{table}[!hbt]
  \tbl{\label{mgozdz:tab2}Contribution to the Majorana neutrino transition
    magnetic moments coming from the bilinear neutrino-neutralino mixing,
    for two GUT scenarios.}{
\begin{tabular}[c]{lcccc}
\toprule
       &  $\mu_{e\mu}$        & $\mu_{e\tau}$      & $\mu_{\mu\tau}$ & trilinear only\\
\colrule
\multicolumn{5}{c}{$A_0=100$, $m_0=m_{1/2}=150\ \GeV$, $\tan\beta=19$}\\
\colrule
IH-CP  &  $3.0\times 10^{-21}$&$2.9\times 10^{-21}$&$2.5\times 10^{-19}$&$\le 10^{-18}$\\
IH-max &  $3.7\times 10^{-19}$&$3.6\times 10^{-19}$&$2.7\times 10^{-19}$&$\le 10^{-18}$\\
NH-CP  &  $3.2\times 10^{-20}$&$3.1\times 10^{-20}$&$2.2\times 10^{-19}$&$\le 10^{-18}$\\
NH-max &  $1.2\times 10^{-19}$&$1.1\times 10^{-19}$&$2.9\times 10^{-19}$&$\le 10^{-18}$\\
\colrule
\multicolumn{5}{c}{$A_0=500$, $m_0=m_{1/2}=1000\ \GeV$, $\tan\beta=19$}\\
\colrule
IH-CP  &  $3.7\times 10^{-22}$&$3.7\times 10^{-22}$&$3.3\times 10^{-20}$&$\le 10^{-20}$\\
IH-max &  $4.6\times 10^{-20}$&$4.6\times 10^{-20}$&$3.5\times 10^{-20}$&$\le 10^{-20}$\\
NH-CP  &  $4.0\times 10^{-21}$&$4.0\times 10^{-21}$&$2.9\times 10^{-21}$&$\le 10^{-20}$\\
NH-max &  $1.4\times 10^{-20}$&$1.5\times 10^{-20}$&$3.7\times 10^{-20}$&$\le 10^{-20}$\\
\botrule
\end{tabular}}
\end{table}
%
The results are presented for two GUT scenarios in
Tab.~\ref{mgozdz:tab2}. The last column shows the upper bounds coming
from pure trilinear RpV coupling constants, ie. from quark-squark and
lepton-slepton loops without bilinear mixing on the external lines. The
conclusion is clear, that the discussed contribution to the main process
is at best of the same order of magnitude, in most cases being at least
an order of magnitude weaker.


\noindent {\bf Acknowledgements. } 
The first author (MG) greatly acknowledges financial support from the
Polish State Committee for Scientific Research under grant no.
N~N202~0764~33.



\end{document}